\documentclass[leqno]{article}
\usepackage{fullpage,subeqn}

\newcommand\be{\begin{equation}}
\newcommand\ee{\end{equation}}
\newcommand\bea{\begin{eqnarray}}
\newcommand\eea{\end{eqnarray}}
\newcommand\nn{\nonumber}
\def\eps{\epsilon}
\def\hzn1{(1 +\epsilon^2 z)^{-1}}

\begin{document}

\bibliographystyle{plain}

\title{Approximate Equations for Large Scale Atmospheric Motions\footnote{Institute 
for Mathematics and Mechanics Report, NYU, May 1951. Retyped June 2006 in \LaTeX  
$\ $ with typos corrected.}}

\author{Joseph B. Keller\footnote{Departments of Mathematics and Mechanical 
Engineering, Stanford University, Stanford, CA  94305-2125, USA}
$\ $ and $\ $
Lu Ting\footnote{Courant Institute of Mathematical Sciences, 
New York University, New York, NY  10012-1185, USA} \\
\\
Institute for Mathematics and Mechanics\\
New York University}

\date{}
\maketitle





\section{Introduction.}
\label{sec:intro}

Recently some meteorologists have attempted to obtain a mathematical 
description of  those large scale atmospheric motions called long waves. 
The primary difficulty is that the equations of gas dynamics, 
which govern the motion of the atmosphere, are so complicated that 
they have not been satisfactorily solved analytically, 
even when viscosity, heat-conduction and moisture are neglected.
They are also unsatisfactory for numerical solution because of the extremely
short time intervals which they necessitate.

In attempting to simplify the equations, meteorologists have often
observed that omission of all acceleration terms from the equations of motion
lead to the hydrostatic pressure and geostrophic wind equations.
The first of these results is considered to be in extremely close agreement
with observation, while the second result is in fairly close agreement 
for large scale motions, particularly at high altitudes where topographical
effects are unimportant.

Although such a derivation of these results is logically somewhat unsatifactory,
a more serious difficulty arises from the attempt to combine them with the 
remaining equations (the conservation of mass and constancy of 
entropy equations). When the hydrostatic pressure and geostrophic wind 
are combined with these equations, it is found that the pressure at the ground
is essentially independent of time.

An atempt to overcome this difficulty has been made by J. Charney. By using the observed values of all the quantities entering the equations, he computes the
magnitude of each term in the equations. He then retains the largest terms
in each equation. In this way he finds that the acceleration terms in the 
equations of motion are small, and thus obtains the hydrostatic pressure and
geostrophic wind. In the mass equation the largest terms are the two which 
constitute the horizontal divergence of the horizontal wind, but the 
vanishing of this divergence does not yield a new result since i
it follows identically
from the geostrophic wind equations. Therefore, another equation 
is obtained
by differentiating and combining the horizontal equations of motion, 
retaining previously neglected terms, and eliminating the horizontal
divergence by means of the mass equation. The largest terms in the resulting
equation are then retained, as well as the complete entropy equation. 
In this way
a semi-empirical deduction of the hydrostatic pressure and geostrophic wind
equation is given, and a complete set of equations embodying them 
is obtained. 
To further simplify the equations additional assumptions, such as that the
wind is independent of height, are made.

Nevertheless it still seemed to us that a systematic mathematical derivation
of the hydrostatic pressure and geostrophic wind equations, 
together with simplified mass and entropy equations , would be worthwhile. 
A method which has been used to derive the shallow water theory, 
the membrane theory of plates, and the theory of thin heavy jets 
immediately suggested itself.

This method involves two steps. First dimensionless variables are introduced 
which involve a small parameter that stretches some coordinates and 
compresses others. The parameter may represent the ratio of s typical vertical
dimension to a typical horizontal dimension of the problem.
Then it is assumed that the solution can be expanded as a power series in 
this parameter. The expansions are inserted into the equations 
and coefficients of
each power of the parameter are equated to zero, yielding a sequence of 
equations for the successive terms in the solution. If an appropriate choice of
dimensionless variables has been made, the first terms 
in the solution satisfy equations of the expected type.
We were guided by Charney's numerical estimates in selecting
our  dimensionless variables.

The result is a simplified set of equations for the
first terms in the solutiom, embodying the hydrostatic
pressure and geostrophic wind equations. This set does
not suffer from the old difficulty of yielding a time-independent
pressure at the ground. It is also simpler than the original set, 
and may yield to approximate solutions. In fact some appromimate solutions 
are given in Section {\bf \ref{sec:special}}. 
Furthermore these simplified equations are 
more suitable for numerical solution than the original equations.

The primary advantage of the present method of derivation is that the dervation
of equations, including the higher order equations, is completely 
automatic once the change of variables has been made.
Thus our equations are slightly different, and in fact simpler, than Charney's
because the expansion scheme determines, for example, 
that a particular coefficient should be a known zero order
quantity rather a sum of known and higher order 
unknown quantities. A less systematic procedure may not yield such results 
because the order of magnitude of every term is not noted.
A secondary advantage of the method is that the mathematical nature of the 
approximation, and its asymptotic character, can be understood, thus leading to an interpretation of the accompanying boundary layer phenomena.

\bigskip

\section{Exact Formulation.}
\label{sec:exact}

We consider the motion of a non-viscous, non-heat conducting,
polytropic gas around the earth. The equations of motion are written in 
Eulerian form employing spherical coordinates referred to a rotating
cordinate system. The axis of rotation is taken to be the polar axis.
The coordinates are radius $r$, colatitude $\theta$ and longitude $\phi$,
and $u, v, w$ are the respective velocity components. The pressure is $p$,
the density $\rho$, and the angular velocity of the coordinate system is
$\Omega$, which is the angular velocity of the earth.
The only external force is that of gravity which 
has the components $-G_1, -G_2$ and $0$ in the  $r, \theta, \phi$ directions. 
The presence of the $\theta$ component is due to the non spherical shape of
the earth and the non-symmetrical mass distribution. The surface of the earth
is given by $r = R(\theta, \phi)$ and the velocity of the gas is assumed to
be tangential to the earth at its surface. In addition $p$ and $\rho$ 
are assumed to approach zero as $r$ becomes infinite.

With these definitions, the three equations of motion
and the equations of conservation of mass and 
constancy of entropy for each ``particle'' become:
\bea
& & u_t + u u_r + v r^{-1} u_{\theta} + 
(r\, \sin \theta)^{-1} w  u_{\phi} = \\
& & r^{-1} (u^2 + w^2) + 2 \Omega w \sin\theta - \Omega^2 r \sin^2\theta 
-G_1 - \rho^{-1} p_r \nn \\
& & v_t + u v_r + v r^{-1} v_{\theta} 
+ (r \sin\theta)^{-1} w v_{\phi} \\
& & = -r^{-1} u v + r^{-1} w^2 \cot \theta + 2 \Omega w \cos \theta
+\Omega^2 r \sin \theta \cos \theta - G_2 - \rho^{-1} p_{\theta} r ^{-1} \nn\\
& & w_t + uw_r + v r^{-1}w_{\theta} + (r \sin \theta)^{-1} w w_{\phi} \\
& & = -r^{-1} uw - r^{-1}vw\cot \theta - 2 \Omega u \sin \theta -
 2\Omega v \cos \theta - (\rho r \sin \theta )^{-1} p_{\phi}\nn \\
& & \rho_t +(\rho u)_r + 
(r \sin \theta)^{-1}(\rho v \sin \theta)_{\theta} 
+(r\sin\theta)^{-1} (\rho w)_{\phi} + \frac{2 \rho u}{r} = 0\\
& & p_t + u p_r + v r^{-1} p_{\theta} + (r \sin \theta)^{-1} w p_{\phi} =
\frac{\gamma p}{\rho} [ \rho_t +  u\rho_r + vr^{-1}\rho_{\theta} 
    + (r \sin \theta )^{-1} w \rho_{\phi}]
\eea
The boundary condition at the earth's surface is:
\be
u -r^{-1} v R_{\theta} - (r \sin \theta )^{-1} w R_{\phi} = 0,
\quad \hbox{at} \ \ r = R(\theta, \phi).
\ee
Equations (1)--(5) are five equations for the five functions
$u, v, w, p, \rho$ assuming $G_1, G_2$ and $R$ are known, In addition
the initial values of the five unknown functions are assumed to be given.

It is convenient to introduce the ``effective'' components
of gravity, defined by:
\bea
g_1 & =& G_1 + \Omega^2 r \sin^2 \theta \\
g_2 & = & G_2 - \Omega^2 r \sin \theta \cos \theta. 
\eea

The formation of the earth is such that at  the surface $r = R(\theta,
\phi)$, the tangential component of ``effective'' gravity is nearly zero,
Since the earth is almost sphereical, this component is practically $g_2$, 
which is consequently small and usually neglected in meteorology.

\section{Dimensionless Variables.}
\label{sec:scaling}

It is convenient to introduce the new independent variable $z = r -a, \ a$
denoting the mean radius of the earth. Then the surface of the earth is given by
$ z = Z(\theta, \phi) \equiv R(\theta , \phi) -a$.

We now introduce dimensionless variables by means of the equations:

\bea
& & \bar u = \epsilon^2 c_0 u, \ \qquad \ \
\bar z = \epsilon^2 a z, \quad \ \ \qquad
\bar t = \epsilon^{-1} a c^{-1}_0 t,
\qquad\ \ \bar Z = \epsilon^4 a Z, \nn \\
& & \bar v = \epsilon^2 c_0 v, \ \ \  \qquad \bar \theta 
= \theta_0 + \epsilon \theta,
\qquad\quad  \bar p = p_0 p, \qquad\qquad\    c_0^2 =  p_0\rho_0^{-1}, \\
& & \bar w = \epsilon^2 c_0 w, \qquad \ \  \bar \phi = \phi_0 + \epsilon \phi,
\qquad \ \ \bar \rho = \rho_0 \rho \qquad \qquad \ \  \mu = \Omega a c_0^{-1}. \nn
\eea
Here the barred quantities are the old variables and 
the unbarred quantities are the corresponding dimensionless variables.
The quantities $\theta_0, \phi_0, c_0, p_0 $ and $\rho_o$ are constants; $c_0$
is a velocity and $p_0, \rho_0$ are typical pressure and density values.
The quatity $\epsilon$ is a small dimensionless parameter which introduces a
stretching in the scale of some quantities and  a contraction in the scale of 
the others. The quantity $\epsilon^2$ may be considered to represent the ratio 
of a typical vertical dimension of the atmosphere to the radius of the earth,
and it is therefore very small. This small parameter will later provide the
basis for a series expansion of the solution.

We also introduce the dimensionless components of ``effective'' gravity, 
$\lambda_1$ and $\lambda_2$ by the equations
\be
g_1ac_0^{-2} = \epsilon^{-3} \lambda_1,
\qquad\qquad g_2 a c_0^{-2} = \epsilon^{\alpha} \lambda_2 .
\ee
The factor $\epsilon^{\alpha}$ makes the smallness of $g_2$ apparent, since
we assume that $\alpha$ is greater than $3$, but is
otherwise unspecified.

Now, introducing equations 7-10 into equations 1-6 we have:
\bea
& & \epsilon^5 u_t + \epsilon^4 u u_z + \epsilon^5 vu_{\theta} 
(1+\epsilon^2 z)^{-1} + \epsilon^5 w u_{\phi} (1 + \epsilon^2 z)^{-1}
(\sin \bar \theta)^{-1} \\
& & \quad = \epsilon^6 (u^2 + w^2)(1 + \epsilon^2 z)^{-1} 
+ \epsilon^4 2 \mu w \sin \bar \theta - \lambda_1 - \rho^{-1} p_z, \nn \\
& & \epsilon^4 v_t + \epsilon^3 uv_z 
+ \epsilon^4 v v_{\theta} (1+\epsilon^2 z)^{-1}
+\epsilon^4 w v_{\phi} (1 + \epsilon^2 z)^{-1} (\sin \bar \theta)^{-1} \\
& & \quad = - \eps^5 u v (1 + \eps^2 z)^{-1} + \eps^5 w^2 \cot \bar \theta \
\hzn1 + \eps^3 2 \mu w \cos \bar \theta 
- \eps^{\alpha + 1} \lambda_2 - \rho^{-1} p_{\theta} \hzn1, \nn \\
& & \eps^4 w_t + \eps^3 u w_z + \eps^4 v w_{\theta} \hzn1 
+ \eps^4 w w_{\phi} (\sin \bar\theta)^{-1} \hzn1 \\
& & \quad = - \eps^5 u w \hzn1 - \eps^5 v w \cot \bar \theta \hzn1
- \eps^3 2 \mu u \sin \bar \theta -\eps^3 2 \mu v \cos \bar \theta
- p_{\phi} (\rho \sin \bar \theta)^{-1} \hzn1, \nn \\
& & \eps \rho_t + (\rho u)_z + \eps (\rho v \sin \bar \theta )_{\theta}
(\sin \bar \theta)^{-1} \hzn1 
+ \eps (\rho w)_{\phi} (\sin \bar \theta)^{-1} \hzn1
+ \eps^2 2 \rho u \hzn1 = 0, \\
& & \eps (p\rho^{-\gamma})_t + u (p \rho^{-\gamma})_z
+ \eps \hzn1 [ v (p \rho^{-\gamma})_{\theta} + \frac{w}{\sin \bar\theta}
( p \rho^{-\gamma})_{\phi}] = 0, \\
& & u = \eps^2 v Z_{\theta}(1 + \eps^4 Z)^{-1} +\eps^2 w Z_{\phi} (\sin \bar \theta)^{-1} 
(1 + \eps^4 Z)^{-1},
\qquad \hbox{at} \ \ z = \eps^2 Z(\theta, \phi). 
\eea

\section{Power Series Solution.}
\label{sec:power}

To solve equations 11-15 subject to the prescribed conditions, we asume that
$ u, v, w, p$ and $\rho$ can be expressed as power series in $\eps$.
Thus we assume

\bea
u = \sum_{i = 0}^{\infty} \eps^i u^i (\theta, \phi, z , t),
& \qquad  &
v = \sum_{i = 0}^{\infty} \eps^i v^i (\theta, \phi, z , t), \nn \\
w = \sum_{i = 0}^{\infty} \eps^i w^i (\theta, \phi, z , t),
& \qquad  &
p = \sum_{i = 0}^{\infty} \eps^i p^i (\theta, \phi, z , t), \\
\rho = \sum_{i = 0}^{\infty} \eps^i \rho^i (\theta, \phi, z , t).
& \qquad  & \nn
\eea

We now insert equations 17 into equations 11-16 and equate to zero the coefficients of each power of $\eps$. From the coefficients of $\eps^0$ we obtain:
\bea
-p^o_z & = & \lambda_1 \rho^0, \\
p^0_\theta & = & 0 ,\\
p^0_{\phi} & = & 0, \\
(\rho^0 u^0)_z & = & 0,\\
u^0 ( p^0_z - \frac{\gamma p^0}{\rho^0} \rho^0_z ) & = & 0, \\
u^0 = 0,   \qquad &\hbox{at}& \quad z = 0.
\eea

From equations 21 and 23 we find $u^0 = 0$, and from equations 19, 20 we have
$p^0 = p^0(z, t)$. Thus the equations 18-23 are equivalent to

\be
u^0 = 0, \quad p^0 = p^0 (z, t), \quad - p^0_z = \lambda_1 \rho^0.
\ee

From the coefficients of $\eps^1$ 
in equations 11-16 we obtain:

\bea
-p^1_z & = & \lambda_1 \rho^1, \\
p^1_{\theta} & = & 0, \\
p^1_{\phi} & = & 0 \\
\rho^0_t + (\rho^0u^1)_z + \rho^0 v^0_{\theta} + \frac{\rho^0}{\sin \theta_0}
w^0_{\phi} & = & 0, \\
p^0_t + u^1 p^0_z & = & \frac{\gamma p^0}{\rho^0} (\rho^0_t 
+ u^1 \rho_z^0), \\
u^1 = 0, \qquad &\hbox{at} &\quad z = 0.
\eea

From the coefficients of $\eps^2$ we have:

\bea
-p^2_z & = & \lambda_1 \rho^2, \\
p^2_{\theta} & = & 0, \\
p^2_{\phi} & = & 0, \\
\rho^1_t + (\rho^0 u^2 + \rho^1 u^1)_z & + & (\rho^1 v^0_{\theta} + 
\rho^0 v^1_{\theta} + \rho^0 v^0 \cot \theta_0) 
+  \frac{1}{\sin \theta_0} ( \rho^0 w^1_{\phi} + \rho^1 w^0_{\phi}
- \rho^0 w^0_{\phi} \theta \cot \theta_0)  =  0,  \\
p^1_t + u^2p^0_z + u^1 p^1_z & = & \frac{\gamma p^0}{\rho^0}
(\rho^1_t + u^2 \rho^0_z + u^1 \rho^1_z )  +  (\frac{\gamma p^1}{\rho^0}
- \frac{\gamma p^0 \rho^1}{(\rho^0)^2})(\rho^0_t + u^1 \rho^0_z), \\
u^2 = 0, \qquad &\hbox{at} & \quad z = 0.
\eea

From the coefficients of $\eps^3$ we have (from equations 11-13)

\bea
-p^3_z & = & \lambda_1 \rho^3 \\
2 \mu w^0 \cos \theta_0 & = & \frac{1}{\rho^0} p^3_{\theta} \\
-2\mu v^0 \cos \theta_0 & = & \frac{1}{\rho^0 \sin \theta_0} p^3_{\phi} .
\eea

We will not write the remaining third order equations,
since they will involve additional coefficients.
Instead we will consider the coefficients of $\eps^4$ in 
equations 12 and 13, which yield

\bea
& & v^0_t + u^1v^0_z + v^0 v^0_{\theta} + w^0 v^0_{\phi} (\sin \theta_0)^{-1}
=  2 \mu w^1 \cos \theta_0 - 2 \mu w^0 \theta \sin \theta_0
- \frac{1}{\rho^0} p^4_{\theta} - \frac{\rho^1 p^3_{\theta}}{(\rho^0)^2}, \\
& & w^0_t + u^1 w^0_z + v^0 w^0_{\theta} + w^0 w^0_{\phi} (\sin \theta_0)^{-1}
\\
& & \quad = -2 \mu u^1 \sin \theta_0 - 2 \mu v^1 \cos \theta_0
+2 \mu v^0 \theta \sin \theta_0
- \frac{p^4_{\phi}}{\rho^0 \sin \theta_0} + p^3_{\phi} (\rho^1 \sin \theta_0
+ \rho^0 \theta \cos \theta_0 )(\rho^0 \sin \theta_0)^{-2}. \nn
\eea
\section{Consequences of the Equations.}
\label{sec:consequence}

Before atttempting to count equations and unknowns,
we will simplify the equations by deducing some obvious consequences of them.
First, by using equations 38, 39 in equation 28 we obtain
\be
\rho_t^0 + (\rho^0 u^1)_z = 0.
\ee
Now using equation 18 in equation 42 yields
\be
-\lambda_1^{-1} p^0_{zt} + (\rho^0 u^1)_z = 0.
\ee

Integrating with respect to $z$ and applying the boundary condition
$\rho^0 = 0$ at $z = \infty$, we have
\be
p^0_t + p^0_z u^1 = 0.
\ee

From equations 44 and 29 we find
\be
u^1_z = 0.
\ee
Using equations 30 and 45, we finally obtain
\be 
u^1 = 0.
\ee

Then from equation 44, $p^0_t = 0$. Thus
\be
p^0 = p^0(z).
\ee

Now of the 16 quantities, $p^0, p^1, p^2, p^3, p^4, \rho^0, \rho^1, \rho^2,
\rho^3, u^0, u^1, u^2, v^0, v^1, w^0, w^1$, which appear in equations 18-47,
two, $u^0$ and $u^1$, are zero (see equations 24, 46). A third, $p^0$, is
independent of $t$ (by eq. 47) and is therefore determined by the initial data.
Of the remaining 13 quantities, 9 ---$\rho^0, \rho^1, \rho^2, \rho^3, u^2, v^0, v^1, w^0, w^1$ --- are given explicitly in terms of the remaining 4,
$p^1, p^2, p^3, p^4$. Of these 4, $p^4$ automatically drops out when
$v^1$ and $w^1$ are eliminated (see eq. 48).
Furthermore $p^2$ appears only in the equation for
$\rho^2$. Thus if only $p^1$ and $p^3$ can be determined, then
$p^0, p^1, p^3, \rho^0, \rho^1, \rho^3, u^2, v^0, w^0$ will be known.

To obtain equations for the determination of $p^1$ and $p^3$, we first
attempt to eliminate $v^1$ and $w^1$ from eq.~34 by means of eqs.~40 and 41.
To this end we differentiate eq.~41 wih respect to $\phi$ and divide it by
$\sin \theta_0$, differentiate eq.~41 with respect to $\theta$, 
and subtract the second from the first. We then  obtain
\bea
& & 2 \mu \cos \theta_0 (\frac{1}{\sin \theta_0} w^1_{\phi} + v^1_{\theta})
\ = \
( v^0_t + v^0 v^0_{\theta} + \frac{w^0 v^0_{\phi}}{\sin \theta_0})_\phi
\frac{1}{\sin \theta_0} \\
& & - (w^0_t + v^0w^0_{\theta} + \frac{w^0 w^0_{\phi}}{\sin \theta_0})_{\theta}
+ \frac{p^3_{\phi\theta} \theta \cot \theta_0}{\rho^0 \sin \theta_0}
+\frac{p^3_{\phi} \cos 2\theta_0}{\rho^0 \sin^2\theta_0 \cos \theta_0}. \nn
\eea
The expression on the left also appears in equation 34.
It is to be noted that $p^4$ does not occur in equation 48.
Now using eq.~48 in eq.~34, and eliminating some terms by the aid of eqs.~38, 39, we have
\bea
& & \rho_t^1 + (\rho^0 u^2)_z +\frac{\rho^0}{2 \mu \cos \theta_0} [
(v^0_t + v^0 v^0_{\theta} + \frac{w^0 v^0_{\phi}}{\sin \theta_0})_{\phi}
\frac{1}{\sin \theta_0} \\
& & - (w_t^0 + v^0 w^0_{\theta} + \frac{w^0 w^0_{\phi}}{\sin \theta_0})_{\theta}
- 2 p^3_{\phi} (\rho^0 \cos \theta_0)^{-1} ] = 0. \nn
\eea

Equation 49 together with equations 39, 38, 35 and 25 are
five equations involving the six unknown functions $v^0, w^0, u^2, p^3, p^1$.
and $\rho^1$. The only other unused equations involving any 
of these quantities are equations 26 and 27 which simply imply
\be
p^1 = p^1(z, t).
\ee
Thus the above equations alone do not seem adequate for the determination
of the unknown functions. If one attempts to supplement them by 
obtaining equations from the higher order terms in the original equations, 
more unknowns are also introduced. Therefore we instead restrict our 
attention to those solutions for which $p^1_t \equiv 0$, i.~e., we pressume that
in any meteorologically significant solution if $p^1$ is independent 
of $\theta$ and $\phi$, it is also independent of $t$.
Then $p^1$ is determined by the initial data, and by equation 25, so is
$\rho^1$. Thus we are left with the four equations 49, 39, 38 and 35 
for the four unknown functions $v^0, w^0, u^2$ and $p^3$.

Equation 35 becomes, since $p^1_t = \rho^1_t =0$,
\be 
u^2(p^0_z - \frac{\gamma p^0}{\rho^0}\rho^0_z) = 0.
\ee
If the second factor, determined by the initial data, is not zero 
(implying that the zero order solution is non-isentropic) then $u^2 = 0$. 
Equations 49, 39 and 38 then suffice to determine $v^0, w^0$ and $p^3$.

On the other hand, if the second factor in equation 51 is zero,
implying that the zeroth order solution is isentropic, this equation is
useless and we remain with three equations for four unknowns. 
To obtain another equation we equate to zero the coefficient of $\eps^3$ 
in equation 15 and find
\be
p^0u^2(\frac{p^1}{p^0} - \frac{\gamma \rho^1}{\rho^0})_z = -
(p^2_t - \frac{\gamma p^0}{\rho^0} \rho_t^2).
\ee
Now from equations 32 and 33, $p^2$ is independent of $\theta$ and $\phi$.
Therefore we restrict our attention to solutions $p^2$ independent of $t$ on
the basis of the presumption mentioned above. Then by eq.~31 $\rho^2$
is also independent of $t$ and both $p^2$ and $\rho^2$ are determined by the initial data. Equation 52 now becomes
\be
u^2 (\frac{p^1}{p^0} - \frac{\gamma \rho^1}{\rho^0})_z =0.
\ee
 
Here again the second factor may not be zero, implying the solution is 
not isentropic to first order, and then $u^2 = 0$.
Then, as before, equations 49, 39 and 38 suffice for the determination 
of $v^0, w^0$ and $p^3$.
If the second factor is zero, equation 53 is useless and we equate to zero 
the coefficient of $\eps^4$ in equation 15 
to obtain the additional equation
\bea
& & p^3_t (\rho^0)^{-\gamma} - \gamma p^0 \rho^3_t (\rho^0)^{-\gamma -1}
    + v^0 [ p^3_{\theta} (\rho^0)^{-\gamma} - \gamma p^0 (\rho^0)^{-\gamma -1}
    \rho^3_{\theta}] \\
& & \quad + \frac{w^0}{\sin \theta_0} [ p^3_\phi (\rho^0)^{-\gamma} 
- \gamma p^0 (\rho^0)^{-\gamma -1} \rho^3_{\phi}] \nn \\
& & \quad + u^2 [ \frac{p^2}{p^0} - \frac{\gamma \rho^2}{\rho^0}
- \frac{\gamma p^1 \rho^1}{p^0 \rho^0}
+ \frac{\gamma(\gamma +1)}{2}\frac{(\rho^1)^2}{(\rho^0)^2} ]_z = 0. \nn
\eea
Now we have the five equations 49, 39, 38, 37, and 54 for the determination
of $v^0, w^0, p^3, \rho^3$ and $u^2$.

\section{Summary of Results.}
\label{sec:summary}

By introducing a certain transformation of variables involving 
a parameter $\eps$, and by assuming that the solution 
can be expanded in powers of $\eps$, we have
obtained a simplified system of equations for the determination of the first terms in the expansion of the solution.
These simplified equations imply that the pressure is
hydrostatic and the horizontal wind geostrophic (to the order in $\eps$
considered). In the course of the derivation it was found that
$p^1$ and $p^2$ are independent of $\theta$ and $\phi$.
We consequently restricted our attention to solutions in which these quantities
are also independent of $t$, presuming that any other solutions 
are not of meteorological importance.
There are two sets of simplified equations, 
depending upon the degree of isentropy of the initial data. 
These two sets are considered separately below.

\subsection{Nonisentropic Case.}
\label{ssec:non-isentropic}

This case obtains if at least one  of the quantities 
$p^0_z - \frac{\gamma p^0}{\rho^0}\rho^0_z$ and
$(\frac{p^1}{p^0} - \frac{\gamma \rho^1}{\rho^0})_z $ is not zero.
Then $u^0 + \eps u^1 + \eps^2 u^2 = 0$,
$p^0(z) + \eps p^1 (z)+ \eps^2 p^2 (z)$ is given by
the initial data and $\rho^0 (z) + \eps \rho^1 (z) + \eps^2 \rho^2(z)
+ \eps^3 \rho^3 ( \theta, \phi, z, t)$ is determined 
by the hydrostatic equation.
Equations 38, 39 and 49 determine $v^0, w^0$ and $p^3$. These equations involve 
no $z$ derivatives, and if $p =p^3/\rho^0$ is introduced as a new unknown, the coefficients are also independent of $z$. The equations then become, omitting 
the superscript on $v^0$ and $w^0$:
\bea
& & w = (2 \mu \cos \theta_0)^{-1} p_{\theta}, \\
& & v = (- 2 \mu \cos \theta_0 \sin \theta_0)^{-1} p_{\phi}, \\
& & (v_t + v v_{\theta} + w v_{\phi} [\sin \theta_0]^{-1})_{\phi}
[\sin \theta_0]^{-1} \\
& & \quad - (w_t + v w_{\theta} + w w_{\phi} [\sin \phi]^{-1})_{\theta}
- 2 [\cos \theta_0]^{-1} p_{\phi} = 0. \nn 
\eea

\subsection{Isentropic Case.}
\label{ssec:isentropic}

This case obtains if both $p^0_z -\frac{\gamma p^0}{\rho^0} \rho^0_z$
and $(\frac{p^1}{p^0} - \frac{\gamma \rho^1}{\rho^0})_z $ are zero. Then
$u^0 + \eps u^1 = 0$, $p^0(z) + \eps p^1(z) + \eps^2 p^2(z)$ is given by the 
initial data and 
$\rho^0(z) + \eps \rho^1 (z)+ \eps^2 \rho^2(z) 
+ \eps^3 \rho^3 ( \theta, \phi, z, t)$ 
is determined by the hydrostatic equation.
Equations 38, 39, 49 and 54 determine $v^0, w^0, p^3$ and $u^2$.
Omitting superscripts and eliminating $\rho^3$ by means of equation 37, 
these equations become:
\bea
& & w = (2 \mu \rho^0 \cos \theta_0)^{-1} p_{\theta} , \\
& & v = (- 2 \mu \rho^0 \cos \theta_0 \sin \theta_0)^{-1} p_{\phi} , \\
& & 2 \mu \cos \theta_0 (\rho^0)^{-1} (\rho^0 u)_z + (v_t + v v_{\theta}
+ w v_{\phi} [\sin \theta_0]^{-1})_{\phi} [\sin \theta_0]^{-1} \\
& & \quad -(w_t +  v w_{\theta} + w w_{\phi} [\sin \theta_0]^{-1})_{\theta}
- 2 [\rho^0 \cos \theta_0]^{-1} p_{\phi} = 0, \nn \\
& & (p + \frac{\gamma p^0}{\lambda_1 \rho^0} p_z)_t
+ v (p + \frac{\gamma p^0}{\lambda_1 \rho^0} p_z )_{\theta}
+ w [\sin \theta_0]^{-1} (p + \frac{\gamma p^0}{\lambda_1 \rho^0} p_z )_{\phi}\\
& & + u (\rho^0)^{\gamma} ( \frac{p^2}{p^0} - \frac{\gamma \rho^2}{\rho^0}
- \frac{\gamma p^1 \rho^1}{p^0 \rho^0}
+\frac{\gamma(\gamma +1)}{2} \frac{[\rho^1]^2}{[\rho^0]^2})_z = 0. \nn
\eea
A simplification of these equations results if the coefficient of $u$ in 
equation 61 is zero, which may be called the extreme isentropic case. Then 
equations 58, 59 and 61 can be solved for $p, w$ and $v$ 
and then $u$ can be found from equation 60.

\section{Boundary Layer Effect.}
\label{sec:boundary}

It may be noticed that the initial data must satisfy various conditions, i.~e., 
geostrophic and hydrostatic equations. Similar conditions must be satisfied
by the boundary data on the spacial boundaries. Stated otherwise, all the
initial and boundary data cannot be prescribed arbitrarily, as one would have 
expected. This is typical of the boundary layer phenomenon which always 
arises in the asymptotic expansion of the solution of a system of 
differential equations, because of the reduced order of the approximate system.
The question arises as to the proper choice of data for the approximate 
solution, when the data for the exact problem are given, in order that 
the approximate solution best approximate the exact solution away from the boundaries. 
This difficult question should not be important in the present case,
however, since the boundaries are not ``real'' but are within a larger region in
which the asymptotic solution is presumably valid.
Therefore the initial and boundary data, if obtained from observations, 
should satisfy the required conditions.

\section{The Barotropic Atmosphere.}
\label{sec:baro}

If the atmosphere is barotropic, i.~e., if there is a functional relation 
between $p$ and $\rho$, then this relation replaces the entropy equation, eq,~5

\vspace{.25in}

\noindent (5') \hspace{2in} $p = f(\rho)$.

\vspace{.25in}

\noindent To derive the simplified equations in this case, we proceed 
exactly as before, but replace all consequences of eq.~5 by those of eq.~5'.
Thus instead of eqs.~22, 29 and 35 we have

\vspace{.25in}

\noindent (22') \hspace{2in} $p^0 = f(\rho^0)$,

\vspace{.25in}

\noindent (29') \hspace{2in} $p^1 = f'(\rho^0) \rho^1$,

\vspace{.25in}

\noindent (35') \hspace{2in} $p^2 = \frac{f''(\rho^0)}{2}[\rho^1]^2
+ f'(\rho^0) \rho^2$.

\vspace{.25in}

\noindent The derivation of eq.~44 is the same as before, but  to proceed
further we restrict our attention to solutions $p^0$ independent of $t$, 
since $p^0$ is already independent of $\theta$ and $\phi$ by eqs.~19 and 20.
Then from eq.~44 we find that $u^1 = 0$.
We further consider only solutions such that $p^1$ and $p^2$ are independent of 
$t$, since both are independent of $\theta$ and $\phi$. 
Equations 38, 39 and 49 follow as before for the determination of
$v^0, w^0, p^3$ and $u^2$.

From equations 38 and 39 we find that $v^0$ and $w^0$ are independent of $z$.
To show this, we differentiate eq.~38 with repect to $z$:
\be
(2 \mu \cos \theta_0) w^0_z = 
\frac{p^3_{\theta z} \rho^0 - p^3_{\theta} \rho^0_z}{(\rho^0)^2}=
\frac{p^0_z \rho^3_{\theta} - p^3_{\theta} \rho^0_z}{(\rho^0)^2}.
\ee
The last equality follows from eqs.~18 and 37. Now from eq.~5' we have
\be
p_{\theta} \rho_z - p_z \rho_{\theta} = 0.
\ee
Since $p^0, p^1$ and $p^2$ are independent of $\theta$ and $\phi$, the
lowest order term in eq.~63 is $p^0_z \rho^3_{\theta} -p^3_{\theta} \rho^0_z$
which is consequently zero. Thus from eq.~62, $w^0_z = 0$ and similarly
$v^0_z = 0$.

Making use of these results, we can eliminate $u^2$ fron eq.~49 by integrating
that equation with respect to $z$ from 0 to $z$, obtaining
\bea
\rho^0 u^2 & = & \{ 
(v^0_t + v^0 v^0_{\theta} 
+ \frac{w^0 v^0_{\phi}}{\sin \theta_0})_{\phi} \frac{1}{\sin \theta_0} - \\
& & \quad (w^0_t + v^0 w^0_{\theta} 
+ \frac{w^0 w^0_{\phi}}{\sin \theta_0})_{\theta} - 
2 p^3_{\phi}[\rho^0\cos \theta_0]^{-1}  \}
\frac{1}{2 \mu \lambda_1 \cos \theta_0}[p^0(z) - p^0(0)].
\nn
\eea
In eq. 64 we have made use of eqs.~18 and 36. Now at $z= \infty$ 
the left side vanishes, and since  $ p^0(\infty)$ also vanishes while $p^0(0)$
is positive, the expression in braces on the right must vanish. This is just 
equation 57. Since this expression is independent of $z$, we find from eq.~64 that $u^2 = 0$. Therefore the quantities $v^0, w^0$ and $p^3$ are determined by
eqs.~55, 56 and 57 in the barotropic atmosphere, just as in the non-isentropic
case (Subsection {\bf 6.1}) for a baroclinic atmosphere.
This result is somewhat surprising, since one might expect the barotropic atmosphere to correspond to the isentropic case. This is also the case, however, for 
if one assumes that the atmosphere is exactly isentropic 
(or at least is up to third order) then eqs.~58 and 59 imply that $v^0$ and 
$w^0$
are independent of $z$; eq.~60 then yields eq.~57 and the result $u^2=0$
as above, and eq.~61 becomes an identity.

\section{Special Solutions (Nonisentropic Case).}
\label{sec:special}

\subsection{Zonal Motion: $v = p_{\phi} = 0$}
\label{ssec:zonal}

Equations 55 and 57 yield

\bea
& & p_{\theta \theta t} = 0. \nn
\eea
Thus, with $a, b, c$ arbitrary functions, we have
\bea 
p & = & a(t) \theta + b (t) + c (\theta) , \nn \\
w & = & (2 \mu \cos \theta_0)^{-1} [ a(t) + c'(\theta)]. \nn
\eea

\subsection{Meridional Motion: $w= p_{\theta} =0$}
\label{ssec:meridianal}

Equations 56 and 57 yield

\bea
& & p_{\phi \phi t} + 4 \mu \sin^2\theta_0 p_{\phi} = 0. \nn
\eea
Integrating
\bea 
& & p_{\phi t} + 4 \mu \sin^2 \theta_0 p = a(t). \nn
\eea
Thus we have
\bea
p & = & a_1(t)  + \int_C f(\alpha) 
e^{\alpha \phi - (4 \mu \sin^2 \theta_0 /\alpha)t} d \alpha , \nn \\
v &= & (-2\mu \cos \theta_0 \sin \theta_0)^{-1}
\int_C \alpha f(\alpha)
e^{\alpha \phi - (4 \mu \sin^2 \theta_0/\alpha) t} d \alpha . \nn
\eea

If we impose periodicity in $\phi$, then $\alpha = n$ 
$(n = 0, \pm 1, \pm 2, \cdots)$ and the integral is replaced by a series.

\subsection{Perturbation of Zonal Motion.}
\label{ssec:perturbation}

Assume a solution analytic in a parameter $\eta$ which  yields a steady
zonal motion for $\eta = 0$. The solution may be written:

\bea
p & = & p^0(\theta) + \eta p^1(\theta, \phi, t) + \eta^2 p^2 + \cdots , \nn \\
w & = & w^0(\theta) + \eta p^1_{\theta} (2 \mu \cos \theta_0)^{-1} + \cdots ,\nn
\\
v & = & (- \mu \sin 2 \theta_0 \sin \theta_0)^{-1} p^1_{\phi \phi t}
+ w^0(\theta)\  p^1_{\phi \phi \phi } (-\mu \sin 2\theta_0 \sin^2 \theta_0)^{-1} 
-(2 \mu \cos \theta_0)^{-1} p^1_{\phi \phi t} \nn \\
& & \quad + (\mu \sin 2 \theta_0)^{-1} (w^0 p^1_{\phi})_{\theta} 
- (2 \mu \cos \theta_0 \sin \theta_0)^{-1} 
(w^0 p^1_{\theta \phi})_{\theta} - 2(\cos \theta_0)^{-1} p^1_{\phi} = 0.
\nn
\eea
If $w^0 = 0$ this simplifies still further to 
\bea
& & \frac{1}{\sin^2 \theta_0} p^1_{t \phi \phi} + p^1_{t \theta \theta}
+ 4 \mu p^1_{\phi} = 0. \nn
\eea
\end{document}